# Machine learning of kinetic energy densities with target and feature averaging: better results with fewer training data

Sergei Manzhos[1,a], Johann Lüder[2,3,4], Manabu Ihara[1]

[1] School of Materials and Chemical Technology, Tokyo Institute of Technology, Ookayama 2-12-1, Meguro-ku, Tokyo 152-8552 Japan

[2] Department of Materials and Optoelectronic Science, National Sun Yat-sen University, 80424, No. 70, Lien-Hai Road, Kaohsiung, Taiwan R.O.C.

[3] Center of Crystal Research, National Sun Yat-sen University, 80424, No. 70, Lien-Hai Road, Kaohsiung, Taiwan R.O.C.

[4] Center for Theoretical and Computational Physics, National Sun Yat-Sen University, Kaohsiung 80424, Taiwan

## Abstract

Machine learning of kinetic energy functionals (KEF), in particular kinetic energy density (KED) functionals, has recently attracted attention as a promising way to construct KEFs for orbital-free density functional theory (OF-DFT). Neural networks (NN) and kernel methods including Gaussian process regression (GPR) have been used to learn Kohn-Sham (KS) KED from density-based descriptors derived from KS DFT calculations. The descriptors are typically expressed as functions of different powers and derivatives of the electron density. This can generate large and extremely unevenly distributed datasets, which complicates effective application of machine learning techniques. Very uneven data distributions require many training data points, can cause overfitting, and ultimately lower the quality of a ML KED model. We show that one can produce more accurate ML models from fewer data by working with partially averaged density-dependent variables and KED. Averaging palliates the issue of very uneven data distributions and associated difficulties of sampling, while retaining enough spatial structure necessary for working within the paradigm of KEDF. We use GPR as a function of partially spatially averaged terms of the 4$^{th}$ order gradient expansion and the Kohn-Sham effective potential and obtain accurate and stable (with respect to different random choices of training points) kinetic energy models for Al, Mg, and Si simultaneously from as few as 2000 samples (about 0.3% of the total KS DFT data). In particular, accuracies on the order of 1% in a measure of the quality of energy-volume dependence $B' = \frac{E(V_0-\Delta V)-2E(V_0)+E(V_0+\Delta V)}{(\Delta V/V_0)^2}$ are obtained simultaneously for all three materials.

---

[a] Author to whom correspondence should be addressed. Email: Manzhos.s.aa@m.titech.ac.jp

## 1 Introduction

Orbital-free density functional theory (OF-DFT)[1–3] has the potential to revolutionize the field of computational materials modeling by making large-scale DFT calculations fast and routine. In Kohn-Sham (KS) DFT[4,5] currently dominating ab initio materials modeling, the energy of a system of $N_{el}$ electrons (we neglect spin without loss of generality) is computed as a functional of the electron density $\rho(\mathbf{r})$,

$$E = -\frac{1}{2}\sum_{i=1}^{N_{el}} \int \psi_i^*(\mathbf{r})\Delta\psi_i(\mathbf{r})d\mathbf{r} + \int V_{ion}(\mathbf{r})\rho(\mathbf{r})d\mathbf{r} + \frac{1}{2}\iint \frac{\rho(\mathbf{r})\rho(\mathbf{r}')}{|\mathbf{r}-\mathbf{r}'|}d\mathbf{r}d\mathbf{r}' + E_{XC}[\rho(\mathbf{r})]$$

(1.1)

where the orbitals $\psi_i(\mathbf{r})$ are the solutions of the Kohn-Sham equation

$$-\frac{1}{2}\Delta\psi_i(\mathbf{r}) + V_{eff}[\rho(\mathbf{r})]\psi_i(\mathbf{r}) = \epsilon_i\psi_i(\mathbf{r})$$

$$V_{eff} = V_{ion}(\mathbf{r}) + \int \frac{\rho(\mathbf{r}')}{|\mathbf{r}-\mathbf{r}'|}d\mathbf{r}' + V_{XC}[\rho(\mathbf{r})]$$

(1.2)

We use atomic units unless stated otherwise. Here $V_{ion}(\boldsymbol{r})$ is the potential due to atomic nuclei, $E_{XC}[\rho(\boldsymbol{r})]$ is the exchange-correlation energy, and $V_{XC}[\rho(\boldsymbol{r})] = \frac{\delta E_{XC}[\rho(\boldsymbol{r})]}{\delta \rho(\boldsymbol{r})}$ is the exchange-correlation potential. The inclusion of $V_{XC}$ into the effective potential $V_{eff}$ is to make sure that $\sum_{i=1}^{N_{el}} |\psi_i(\boldsymbol{r})|^2 = \rho(\boldsymbol{r})$, the true electron density.[5] The need to compute the orbitals causes a near-cubic scaling of the computational cost with system size, which is further exacerbated by the need to ensure self-consistent convergence of orbital-dependent $V_{eff}\left[\rho(\boldsymbol{r}) = \sum_{i=1}^{N_{el}} |\psi_i(\boldsymbol{r})|^2\right]$, as the Eq. (1.2) is typically solved as a linear ODE for a given $\rho(\boldsymbol{r})$. As a result, routinely doable calculations are limited to $10^2$-$10^3$ atoms. While linear scaling approaches to KS-DFT exist,[6–9] typically based on the use of finite-support basis functions to represent $\psi_i(\boldsymbol{r})$, the linear scaling is only achieved in the limit of large systems (more than dozens of Å) with substantial CPU cost necessitating the use of supercomputers.

In Eq. (1.1), only the first term, which is the Kohn-Sham kinetic energy

$$E_{kin}^{KS} = -\frac{1}{2} \sum_{i=1}^{N_{el}} \int \psi_i^*(\boldsymbol{r}) \Delta \psi_i(\boldsymbol{r}) d\boldsymbol{r} \equiv \int \tau_{KS}(\boldsymbol{r}) d\boldsymbol{r}$$

(1.3)

requires the orbitals explicitly (we do not consider here approaches such as hybrid functionals where $E_{XC}$ is made explicitly depend on orbitals). Here, we defined the Kohn-Sham kinetic energy density (KED) $\tau_{KS}(\boldsymbol{r})$. It is sometimes convenient to use a positively definite KED $\tau_+(\boldsymbol{r})$,

$$\tau_+(\boldsymbol{r}) = \frac{1}{2} \sum_{i=1}^{N_{el}} |\nabla \psi_i(\boldsymbol{r})|^2 = \tau_{KS}(\boldsymbol{r}) + \frac{1}{4} \Delta \rho(\boldsymbol{r})$$

(1.4)

that integrates to the same $E_{kin}^{KS}$. Using $E_{kin} = E_{kin}[\rho(\boldsymbol{r})] \approx E_{kin}^{KS}[\rho(\boldsymbol{r})]$ without explicit dependence on $\psi_i(\boldsymbol{r})$ gives rise to OF-DFT whereby one optimizes the density directly to minimize the total energy. $E_{kin}[\rho(r)]$ is the kinetic energy functional (KEF). Main types of approximations being developed for $E_{kin}[\rho(r)]$, non-local and semi-local, scale much better with system size than KS DFT. In the non-local approach, one typically uses[10–15]

$$E_{kin}[\rho(r)] = E_{TF}[\rho(r)] + E_{vW}[\rho(r)] + E_{NL}[\rho(r)]$$

(1.5)

where $E_{TF}[\rho(r)] = \frac{3}{10}(3\pi^2)^{2/3} \int \rho^{5/3}(r)dr$ is the Thomas-Fermi (TF) kinetic energy[16] and $E_{vW}[\rho(r)] = \frac{1}{8}\int \frac{|\nabla\rho(r)|^2}{\rho(r)} dr$ is the von Weizsäcker (vW) kinetic energy.[17] The non-local term $E_{NL}[\rho(r)]$ is expressed as

$$E_{NL}[\rho(r)] = \iint \rho^\alpha(r)\omega(r,r')\rho^\gamma(r')drdr'$$

(1.6)

which in general results in a quadratic to $n\ log(n)$ scaling. In the semi-local approach, one models the KED and the corresponding KEF are therefore kinetic energy density functionals (KEDF):

$$E_{kin}[\rho(r)] = \int \tau[\rho(r)]dr$$

(1.7)

where one typically strives $\tau[\rho(r)] \approx \tau_{KS}(r)$ or $\tau[\rho(r)] \approx \tau_+(r)$. Here dependence on $\rho(r)$ means dependence on any quantities dependent on the density and its derivatives at point $r$. The semi-local approach is more apt to achieve near-linear scaling.

As a result, (with both non-local and semi-local KEFs) systems with tens of thousands of atoms are computable on a desktop computer, with millions of atoms routinely computable with access to supercomputers.[18,19] This opens a door to direct DFT modeling of intrinsically large-scale phenomena such as microstructure-driven properties, disordered systems etc.[19–26] Unfortunately, existing KEF approximations are only accurate enough for use in real-life applications for light metals.[19,21,27–32] The development of accurate KEFs for other classes of materials remains a significant bottleneck on the way to wider use of OF-DFT in applications.

Recently, machine learning (ML) has been increasingly used for the construction of KEFs, in particular KEDFs.[33–44] Neural network (NN)[45] and kernel-based[46] methods such as kernel ridge regression (KRR) and Gaussian process regression (GPR) [47] have been mostly used. These widely used methods will not be introduced here beyond the aspects important for this work; the reader is referred to the above references for their description. These two classes of methods have their advantages and disadvantages, as we recently highlighted in Ref. [48]. In particular, when data are abundant, NNs have a higher expressive power than kernel methods that are essentially linear regressions, while the latter are, for that reason, more robust with respect to overfitting and may provide better results when density of sampling is low.[49] The density of sampling is expected to be a limiting factor when the dimensionality of the feature space is high and the data distribution is very uneven, as is in fact the case when fitting KED. The distributions of the KED itself as well as of various typically used density-dependent features such as different derivatives and powers of the density are very uneven, as can be appreciated e.g. from Fig. 5 of Ref. [35] as well as from the figures below. Prima faci this calls for a large number of training samples. This may be complicated in particular with GPR that has to wield the inverse of a matrix of size $M \times M$ where $M$ is the number of training points. This becomes costly when $M$ exceeds about $10^4$. While various workarounds exist that facilitate using GPR with large training sets,[50] it would be advantageous to have a way to fit KED *or a quantity that can be used in the same way as the KED* (i.e. integrating to the same kinetic energy) that allows reliable machine learning from small training sets. We show here that this is possible.

A key test of the potential of a machine-learned KEF to be usable in applications is its ability to be used for structure optimization,[43,44,51] which is not guaranteed just by a good fit to $\tau_{KS,+}(\mathbf{r})$. Machine-learned functionals can result in stable structure optimization.[33,35,36] In particular, in Ref. [35], when machine-learning KEDs of Mg, Al, and Si with GPR, we considered the ability of the GPR model to reproduce the energy-volume dependence, which is a measure of ability and quality of optimization expected with a given KED model. While a qualitatively accurate energy-volume dependence may suffice for structure optimization, quantitative accuracy of the said dependence would be needed to compute phononic properties.

In this work, we also consider GPR of KED of Mg, Al, and Si and monitor the quality of the energy-volume dependence. *We specifically focus on obtaining reliable ML models of the KE even with small training sets sampling from large, extremely unevenly distributed KS DFT-derived datasets*. We show that it is possible to obtain high quality KE models, resulting in accurate energy-volume dependence, with rather small datasets, by machine learning from partially spatially averaged density-dependent variables and a partially averaged KED that integrates to the same kinetic energy. Averaging palliates the issue of very uneven data distributions of the features and of the kinetic energy density. Because the averaging is partial (i.e. smoothing), we preserve spatial information which is necessary for working within the paradigm of KEDF although one machine-learns not a real KED but a smoother space-dependent function that integrates to the same kinetic energy. We use GPR as a function of partially spatially averaged terms of the 4$^{th}$ order gradient expansion[52] and the product of the density and the KS effective potential and obtain accurate and stable (with respect to the choice of particular random samples) kinetic energy models from as few as 2000 samples (about 0.3% of the total KS DFT data). In particular accuracies on the order of 1% in a measure of the quality of energy-volume dependence $B' = \frac{E(V_0-\Delta V)-2E(V_0)+E(V_0+\Delta V)}{(\Delta V/V_0)^2}$ are obtained simultaneously for Al, Mg, and Si.

## 2 Methods

We use similar reference KS DFT data as in Ref. [35]. Briefly, calculations on Mg, Al, and Si were performed in Abinit[53,54] using PBE exchange-correlation functional[55] and a plane-wave cutoff of 500 eV. Real space local pseudopotentials from Carter's group were used.[29] We used local pseudopotentials a those are most likely to be used in OF-DFT calculations (as nonlocal pseudopotentials cannot be used in OF-DFT directly[56–58]). Total energies were converged to 1×10$^{-7}$ *a.u.*, and structures were optimized until all force components were below 1×10$^{-4}$ *a.u.* Conventional standard unit cells were used for face-centered cubic Al, cubic diamond Si, and hexagonal closed packed cell for Mg. The Brillouin zone was sampled with 8×8×8, 6×6×6, and 10×10×8 $k$-points, respectively. Optimal lattice constant were 4.047 Å for Al and 5.470 Å for Si while for Mg $a$ = 3.18 Å, $c$ = 5.25 Å, in good agreement with literature.[59] Electron densities, positively definite KEDs ($\tau_+$), gradients and Laplacians of the density, and Kohn-Sham effective potentials were output from Abinit on the entire Fourier grid (the real space grid equivalent to the plane wave cutoff energy). Data were collected at the equilibrium geometry as well as uniformly compressed or expanded cells with all lattice constants multiplied by 1-δ and 1+ δ (volume changes by (1±δ)³), respectively, with δ=0.05. The data thus include materials with different types of bonding. All these datasets are concatenated and machine-learned together thereby creating an element of portability of the KED model.

We machine-learn $\tau_{KS}(\boldsymbol{r})$ as a function of the following seven density-dependent features:

$$x = (\tau_{TF}, \tau_{TF}p, \tau_{TF}q, \tau_{TF}p^2, \tau_{TF}pq, \tau_{TF}q^2, \rho V_{eff})$$

(2.1)

where $\tau_{TF}$ is the Thomas-Fermi KED, $p = \frac{|\nabla \rho|^2}{4(3\pi^2)^{2/3}\rho^{8/3}}$ is the scaled squared gradient and $q = \frac{\Delta \rho}{4(3\pi^2)^{2/3}\rho^{5/3}}$ is the scaled Laplacian of the density. The scaling helps satisfy the so-called exact conditions.[60] The first six features are the terms of the 4$^{th}$ order gradient expansion[52] $\tau_{GE4} = \tau_{TF}\left(1 + \frac{5}{27}p + \frac{20}{9}q + \frac{8}{81}q^2 - \frac{1}{9}pq + \frac{8}{243}p^2\right)$ that were previously shown to be good density-dependent variables when machine learning KED.[34,35] The term $\rho V_{eff}$ is responsible for a significant fraction of the variance of the KED (as $\tau_{KS}(\boldsymbol{r}) + V_{eff}(\boldsymbol{r})\rho(\boldsymbol{r}) = \sum_i \epsilon_i |\psi_i(\boldsymbol{r})|^2$) and can help regression.[35] This term is orbital-independent unless hybrid functionals or meta functionals with orbitals-dependent KE terms are used.

To evaluate the quality of the energy-volume dependence, similar to Ref. [35], we define a bulk modulus-like quantity

$$B' = V_0^2 \frac{d^2 E}{dV^2} \approx \frac{E(V_0 - \Delta V) - 2E(V_0) + E(V_0 + \Delta V)}{(\Delta V/V_0)^2}$$

(2.2)

where $V_0$ is the equilibrium volume of the simulation cell and $E$ is the total energy. With machine-learned KED models, the predicted total energy $E$ is computed as $E_{pred} = E_{KS} - \int (\tau_{KS} - \tau_{pred}) d\boldsymbol{r}$. This serves to monitor the ability of the KED model to reproduce the energy-volume dependence important for structure optimization. We also define a similar quantity but based on the kinetic energy alone,

$$B'_{kin} = V_0^2 \frac{d^2 E_{kin}}{dV^2} \approx \frac{E_{kin}(V_0 - \Delta V) - 2E_{kin}(V_0) + E_{kin}(V_0 + \Delta V)}{(\Delta V/V_0)^2}$$

(2.3)

We show below that monitoring the quality of the KED fit alone and $B'_{kin}$ can give a deceptive view of the ability of the model for structure optimization, as relative errors are higher for $B'$ than for $B'_{kin}$.

We form partially spatially averaged (smoothened) features and target,

$$\bar{x}_i(\boldsymbol{r}) = \int w(\boldsymbol{r}, \boldsymbol{r}') x_i(\boldsymbol{r}') d\boldsymbol{r}'$$

$$\overline{\tau_{KS}}(\boldsymbol{r}) = \int w(\boldsymbol{r}, \boldsymbol{r}') \tau_{KS}(\boldsymbol{r}') d\boldsymbol{r}'$$

(2.4)

where the integration is over the entire simulation cell, and $w(\boldsymbol{r}, \boldsymbol{r}')$ is a kernel such that $\int w(\boldsymbol{r}, \boldsymbol{r}') d\boldsymbol{r}' = \int w(\boldsymbol{r}', \boldsymbol{r}) d\boldsymbol{r}' = 1, \forall \boldsymbol{r}$, where, in general, $\boldsymbol{r}' \subseteq \boldsymbol{r}$. It is easy to see that the integration of $\overline{\tau_{KS}}$ results in the same kinetic energy: $\int \overline{\tau_{KS}}(\boldsymbol{r}) d\boldsymbol{r} = \int \tau_{KS}(\boldsymbol{r}) [\int w(\boldsymbol{r}, \boldsymbol{r}') d\boldsymbol{r}'] d\boldsymbol{r} = E_{kin}$. The benefit of averaging is alleviating the issue of very uneven distributions of $x_i$ and of the KED that complicates sampling and results in overfitting, higher computational cost, and ultimately lower model accuracy. Because the averaging is partial (smoothing), spatial information is preserved and allows working in the paradigm of KEDF. This is different from fitting in the average sense (averaging the prediction as a linear operation) which is always present due to a finite width of the kernel in GPR; because the features are averaged that enter a non-linear GPR kernel (Eq. (2.5)), the effect of the averaging that we do is non-linear. The original KS DFT and averaged data are available from the authors at reasonable request.

We explored averaging over a cube built with $\pm n_{ave}$ grid points around $r$ thus averaging over about $N_{ave} = (2n_{ave})^3$ grid points, i.e. $w(r, r') = \frac{1}{N_{ave}}$ within the cube and zero elsewhere, and averaging with a Gaussian function $w(r, r') = (\sigma\sqrt{2\pi})^{-3} exp\left(-\frac{|r-r'|^2}{2\sigma^2}\right)$. When these kernels had similar special extents, we did not note significant differences in achievable quality of the model. What is important is the spatial extent of the averaging and not a particular way it is done. In that follows we present results with $w(r, r') = \frac{1}{N_{ave}}$ for different $n_{ave}$.

GPR calculations were done in Matlab using the *fitrgp* function and a Matern32 kernel,

$$k(x_i, x_j) = \sigma^2 \left(1 + \frac{\sqrt{3}|x_i - x_j|}{l}\right) exp\left(-\frac{\sqrt{3}|x_i - x_j|}{l}\right)$$

(2.5)

where $\sigma^2$ is the variance of the target and $l$ is the length hyperparameter. The features $x$ were scaled to a unit cube and therefore an isotropic kernel was used with a single length parameter.

## 3 Results

### 3.1 GPR fitting of original DFT-derived data

We first perform calculations with the original KS DFT-derived data as a reference. We use $M$ = 2000, 5000, and 10000 training points and 20000 test points. A significant component of the cost of GPR is the calculation of the inverse of the covariance matrix between all pairs of the training points,[47] which begins to become difficult beyond about 10000 training points. We therefore limit ourselves to a maximum of 10000 training points (which is also sufficient to get good accuracy of $B'$ as is shown below). The total dataset contains about 585000 data points, so a natural question is whether these relatively small train and test samples are representative. To compute $B'$ and $B'_{kin}$, we call the model on all points and therefore can compare the errors on the entire set to the errors on the (relatively small) test and train sets to answer this question.

Table 1 shows train and test set kinetic energy density RMSE (root mean square error), as well as RMSE over the entire dataset, kinetic energies of Al, Mg, and Si at equilibrium and strained simulation cells, as well as $B'$ and $B'_{kin}$, when fitting Kohn-Sham KED $\tau_{KS}$ using different numbers of training points. We also show in the table the mean relative error (MRE) of $B'$ and $B'_{kin}$ over all materials from the reference Kohn-Sham values, $MRE(B') = \frac{1}{3}\sum_{Al,Mg,Si} abs\left(1 - \frac{B'}{B'_{KS}}\right)$ and $MRE(B'_{kin}) = \frac{1}{3}\sum_{Al,Mg,Si} abs\left(1 - \frac{B'_{kin}}{B'_{kin,KS}}\right)$. GPR was previously used on the same data with 2000 and 5000 training points, using a different code (Octave) and a committee of 5 GPRs, machine learning $\tau_+(r)$, and computing only $B'$ but not $B'_{kin}$, in Ref. [35]. There is overall agreement with those results, which obtained $B'$ of 0.914 (Al), 0.444 (Mg), and 2.005 (Si) with 2000 training points and 1.015 (Al), 0.421 (Mg), and 2.981 (Si) with 5000 training points. The results in the table are obtained with optimized hyperparameters, whereby the length parameter and the regularization parameter were scanned for the best test set errors. With the thus found hyperparameters, we performed ten fits that differed by different random selections of $M$ training points from the total dataset. The table shows ranges of values due to random point selection.

The following conclusions can be made from these results: a 20000-point test set is sufficient to evaluate the KED model quality (i.e. the test RSME is similar to the all-data RMSE). This is non-trivial considering that the test set is a small fraction of the overall dataset and that the distributions of the KED and of the features are very uneven, see Fig. 5 of Ref. [35] and Supplementary Material. There remains a significant spread of values due to the random selection of training points; the size of the test set is however sufficient to account for it, and there is good correlation between the test set RMSE and the total dataset RSME (the Pearson correlation coefficient $R$ of more than 0.8), so that hyperparameters for the best global RMSE can be chosen by monitoring the error on a test set of this size. One can perform several fits (with different random draws of training points) monitoring the test set error and select the best.

Table 1. The RMSE (root mean square error) of the kinetic energy density (KED) for train, test, and total datasets, kinetic energies $E_{kin}$ of Al, Mg, Si at equilibrium and isotopically strained simulation cells (top to bottom: $V_{equil}$, $V_{equil} - \Delta V$, $V_{equil} + \Delta V$), as well as $B'$ and $B'_{kin}$, when fitting Kohn-Sham KED using different numbers of training points (the number of test points is 20000 in all cases). All values are in atomic units. Optimal hyperparameters (length parameter $l$ and logarithms of regularization parameter $\delta$) are also given. Mean relative errors (MRE) from the reference Kohn-Sham values of $B'$ and $B'_{kin}$ over all materials are also given. The ranges of values obtained due to random nature of train point selection are given.

| Train points | $l$ /log($\delta$) | Train RMSE | Test RMSE | All data RMSE | $E_{kin}$ Al | $E_{kin}$ Mg | $E_{kin}$ Si | $B'_{kin}$ / $B'$ Al | $B'_{kin}$ / $B'$ Mg | $B'_{kin}$ / $B'$ Si | MRE ($B'_{kin}$) | MRE ($B'$) |
|---|---|---|---|---|---|---|---|---|---|---|---|---|
| 2000 | 1.4 / -3.5 | 0.0138- 0.0157 | 0.1286- 0.2404 | 0.1349- 0.2341 | 3.3078- 3.3153 / 3.6485- 3.6612 / 3.0381- 3.0483 | 0.6822- 0.6830 / 0.7590- 0.7608 / 0.6193- 0.6211 | 11.7864- 11.8115 / 12.5750- 12.6202 / 11.1871- 11.2035 | 2.661- 3.213 / / / 0.893- 1.444 | 0.517- 0.656 / / / 0.344- 0.483 | 6.633- 8.321 / / / 3.357- 5.045 | 0.069- 0.185 | 0.137- 0.352 |
| 5000 | 1.4 / -4.5 | 0.0023- 0.0026 | 0.0770- 0.1059 | 0.0734- 0.1049 | 3.3107- 3.3147 / 3.6555- 3.6583 / 3.0377- 3.0411 | 0.6819- 0.6827 / 0.7599- 0.7610 / 0.6189- 0.6202 | 11.7939- 11.8063 / 12.5993- 12.6097 / 11.1680- 11.1781 | 2.753- 2.990 / / / 0.984- 1.222 | 0.549- 0.643 / / / 0.375- 0.469 | 6.620- 7.473 / / / 3.344- 4.197 | 0.058- 0.112 | 0.105- 0.204 |

| | | | | | | | | | | | |
|---|---|---|---|---|---|---|---|---|---|---|---|
| 10000 | 1.0 / -4 | 0.0040-0.0043 | 0.0437-0.0722 | 0.0424-0.0866 | 3.3127-3.3144 | 0.6826-0.6830 | 11.8006-11.8036 | 2.774-2.848 | 0.554-0.590 | 6.572-6.910 | 0.031-0.055 | 0.054-0.105 |
| | | | | | 3.6565-3.6585 | 0.7602-0.7606 | 12.6030-12.6094 | / | / | / | | |
| | | | | | 3.0382-3.0401 | 0.6193-0.6197 | 11.1628-11.1663 | 1.006-1.080 | 0.381-0.417 | 3.296-3.634 | | |
| KS reference | | | | $V_0 - \Delta V$ | 3.3134 | 0.6831 | 11.8046 | 2.828 /1.060 | 0.538 /0.365 | 6.368 /3.092 | | |
| | | | | $V_0$ | 3.6585 | 0.7604 | 12.6083 | | | | | |
| | | | | $V_0 + \Delta V$ | 3.0386 | 0.6192 | 11.1591 | | | | | |

Table 2. The RMSE (root mean square error) of the kinetic energy density (KED) for train, test, and total datasets, kinetic energies $E_{kin}$ of Al, Mg, Si at equilibrium and isotopically strained simulation cells (top to bottom: $V_{equil}$, $V_{equil} - \Delta V$, $V_{equil} + \Delta V$), as well as $B'$ and $B'_{kin}$, when fitting partially spatially averaged Kohn-Sham KED with $n_{ave} = 10$ using different numbers of training points (the number of test points is 20000 in all cases). All values are in atomic units. Optimal hyperparameters (length parameter $l$ and logarithms of regularization parameter δ) are also given. Mean relative errors (MRE) from the reference Kohn-Sham values of $B'$ and $B'_{kin}$ over all materials are also given. The ranges of values obtained due to random nature of train point selection are given.

| Train points | $l$ /log(δ) | Train RMSE | Test RMSE | All data RMSE | $E_{kin}$ Al | $E_{kin}$ Mg | $E_{kin}$ Si | $B'_{kin}$ / $B'$ Al | $B'_{kin}$ / $B'$ Mg | $B'_{kin}$ / $B'$ Si | MRE ($B'_{kin}$) | MRE ($B'$) |
|---|---|---|---|---|---|---|---|---|---|---|---|---|
| 2000 | 1.4 / -5.5 | 0.0000- 0.0000 | 0.0033- 0.0059 | 0.0033- 0.0057 | 3.3134- 3.3134 | 0.6832- 0.6832 | 11.8042- 11.8048 | 2.832- 2.839 | 0.529- 0.530 | 6.350- 6.390 | 0.006- 0.008 | 0.010- 0.013 |
|  |  |  |  |  | 3.6585- 3.6586 | 0.7604- 0.7604 | 12.6073- 12.6086 | / | / | / |  |  |
|  |  |  |  |  | 3.0387- 3.0387 | 0.6192- 0.6192 | 11.1590- 11.1599 | 1.064- 1.070 | 0.356- 0.357 | 3.074- 3.114 |  |  |
| 5000 | 1.4 / -5.5 | 0.0000- 0.0000 | 0.0015- 0.0030 | 0.0015- 0.0030 | 3.3134- 3.3134 | 0.6832- 0.6832 | 11.8045- 11.8048 | 2.835- 2.837 | 0.529- 0.530 | 6.376- 6.389 | 0.007- 0.007 | 0.012- 0.012 |
|  |  |  |  |  | 3.6585- 3.6586 | 0.7604- 0.7604 | 12.6080- 12.6085 | / | / | / |  |  |
|  |  |  |  |  | 3.0387- 3.0387 | 0.6192- 0.6192 | 11.1594- 11.1596 | 1.067- 1.069 | 0.356- 0.356 | 3.100- 3.113 |  |  |

| | | | | | | | | | | | |
|---|---|---|---|---|---|---|---|---|---|---|---|
| 10000 | 1.2 / -5 | 0.0000-0.0000 | 0.0008-0.0011 | 0.0009-0.0012 | 3.3134-3.3134 | 0.6832-0.6832 | 11.8046-11.8047 | 2.834-2.836 | 0.529-0.530 | 6.377-6.385 | 0.007-0.007 | 0.011-0.012 |
| | | | | | 3.6585-3.6585 | 0.7604-0.7604 | 12.6084-12.6085 | / | / | / | | |
| | | | | | 3.0387-3.0387 | 0.6192-0.6192 | 11.1594-11.1595 | 1.066-1.067 | 0.356-0.356 | 3.101-3.109 | | |
| KS reference | | | | $V_0 - \Delta V$ | 3.3134 | 0.6831 | 11.8046 | 2.828 /1.060 | 0.538 /0.365 | 6.368 /3.092 | | |
| | | | | $V_0$ | 3.6585 | 0.7604 | 12.6083 | | | | | |
| | | | | $V_0 + \Delta V$ | 3.0386 | 0.6192 | 11.1591 | | | | | |

The apparent good accuracy and small relative scatter (due to different random selections of training points) of the predicted values of $E_{kin}$ to the KS values belies more significant errors and a significant spread of errors in the energy-volume dependence $B'$. The maximum (over multiple possible selections of train points) error trends down with the training set size reaching close to 11% and 6% for $B'$ and $B'_{kin}$, respectively, with 10000 training points. The best error reaches about 5% and 3% respectively, with 10000 training points, i.e. a significant spread due to a particular random selection of training points remains even with $M = 10000$. 5000 training points are sufficient to achieve the best accuracies of $B'$ and $B'_{kin}$ of about 11% and 6%, respectively, and maximum MRE on the order of 20% and 11%, respectively. We find that while there exists a positive correlation between MRE and the RMSE on the full dataset, it is not high, on the order of 0.3. That is, achieving the best KED RMSE even on *all* points entering kinetic energy calculation may not result in the best KED model. This highlights the difficult nature of this application from the perspective of data science, whereby the role of data and error distributions is important. It therefore offers a stringent test for a ML method. Note that the error of the $B'$ is higher than that of $B'_{kin}$ suggesting that the dependence of the total energy on volume should be monitored and not just the error in the kinetic energy.

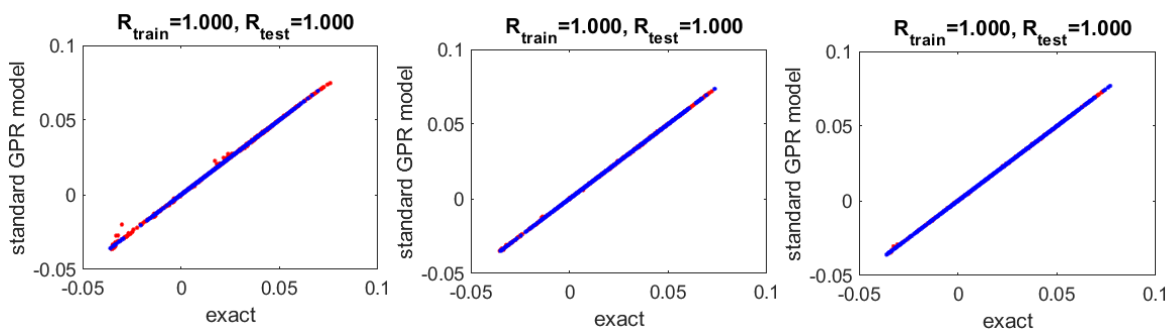

Figure 1. Correlation plots between Kohn-Sham ("exact") KED values (in a.u.) and predicted KED with GPR using different numbers of train points (left to right: 2000, 5000, 10000). Blue: train points, red: test points. Where test points are not visible, they visually overall with the training points.

Representative regression plots between the KS KED and the model are shown in Figure 1. The quality of regression is high with Pearson correlation coefficients $R$ as good as 1.000 for both train and test sets (irrespective of different random draws of training points) with very few outliers from the diagonal trend. There is no visible difference in fit quality between 5000 and 10000 training points.

*3.2 GPR on partially spatially averaged data*

We performed tests with $n_{ave}$ of 5, 10, 15, and 20 grid points. The plane wave cutoff used in the parent KS DFT calculations results in grid spacing of about 0.11-0.12 Å. A constant $n_{ave}$ therefore results in averaging over an approximately constant spatial extent for all systems considered (it is desired for portability that the averaging window be system-independent). Beyond $n_{ave} = 20$, the size of the averaging kernel approaches the simulation cell sizes of some of the systems, this is therefore the largest kernel size we use.

The effect of averaging is shown in Figure 2, where histograms of the KED and of the features are shown, original and partially spatially averaged for $n_{ave} = 10$. The effect of averaging is qualitatively similar for other values of $n_{ave}$. The effect is best viewed on the logarithmic scale, on which Figure 2 is plotted. The non-logarithmic scale version is provided in Supplementary Material; it highlights the extremely uneven distributions of the data, which are spiky with long sparsely sampled tails of large KED or $x_i$ values, which are trimmed by the averaging.

Table 2 shows fitting results for $n_{ave} = 10$ in the same format as Table 1. The RMSE of the train, test, and all-point datasets is significantly improved, to the point where the RMSE on the train set is negligible (zero on the scale of the table, which is the same as in Table 1). Of course, averaging the data makes fitting it easier; it is trivial that train and test set errors decrease. It is also expected that smoother data lead to a lower optimal value of δ. *What is important is not a decrease in train and test RMSE but the fact that the quality of $E_{kin}$ is significantly improved, and the spread of values due to different random selections of train points is significantly diminished* (as the spread is minor when using averaging, 5 random point selections were considered in this test). Even with only 2000 training points, one achieves errors in $B'_{kin}$ and $B'$ that are on the order of 1% vs KS DFT and differ by less than 0.3 % for different draws of training points.

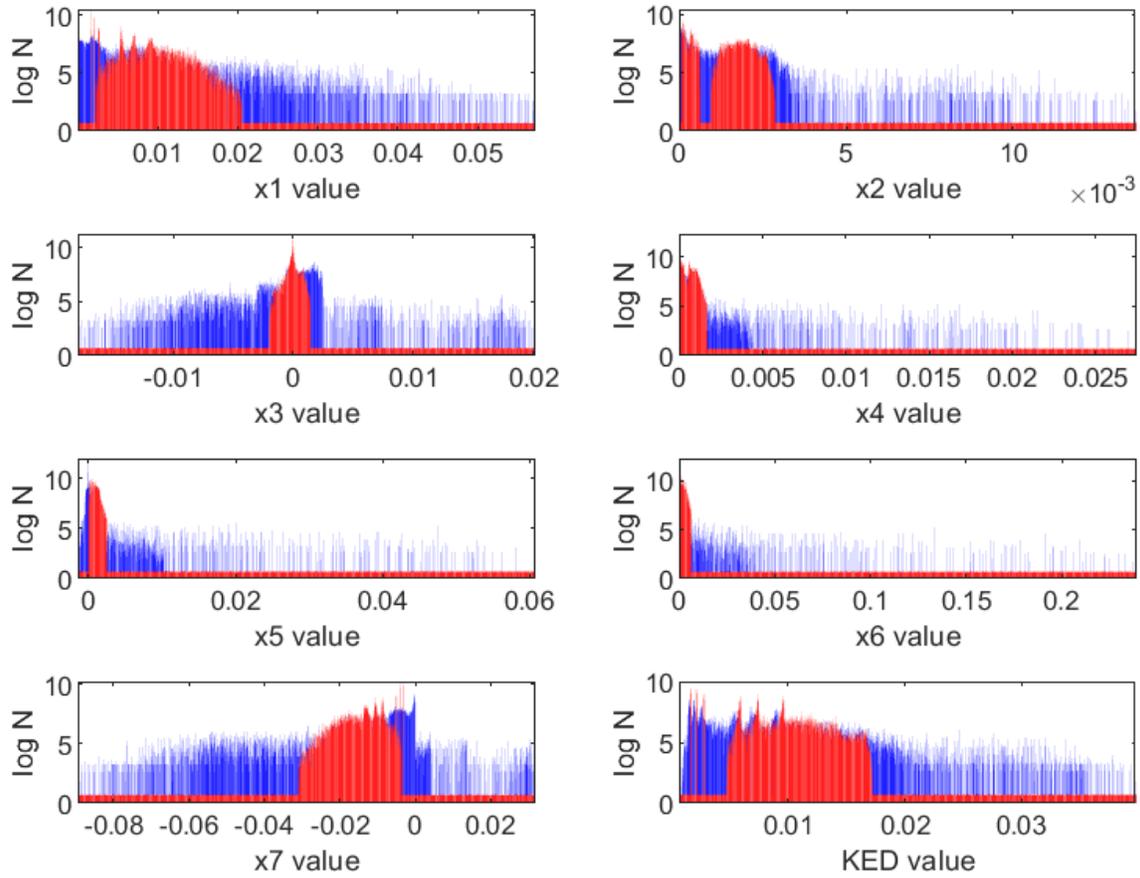

Figure 2. Histograms of features and KED. Blue: original KS DFT grid-based; red: averaged with $n_{ave} = 10$. Note the logarithmic scale (N: number of values in any of 2000 bins).

With 5000 training points, there is no noticeable dependence on a particular random points selection and convergence is achieved of $B'$ and $B'_{kin}$ with accuracies of this quantities of about 1.1% and 0.7%, respectively.

Representative regression plots between the (partially spatially averaged) KS KED values and the model are shown in Figure 3. The quality of regression is high with Pearson correlation coefficients $R$ as good as 1.000 for both train and test sets (irrespective of different random draws of training points) with *no* outliers from the diagonal trend even at 2000 training points (cf. Figure 1). There is no visible difference in fit quality between 5000 and 10000 training points.

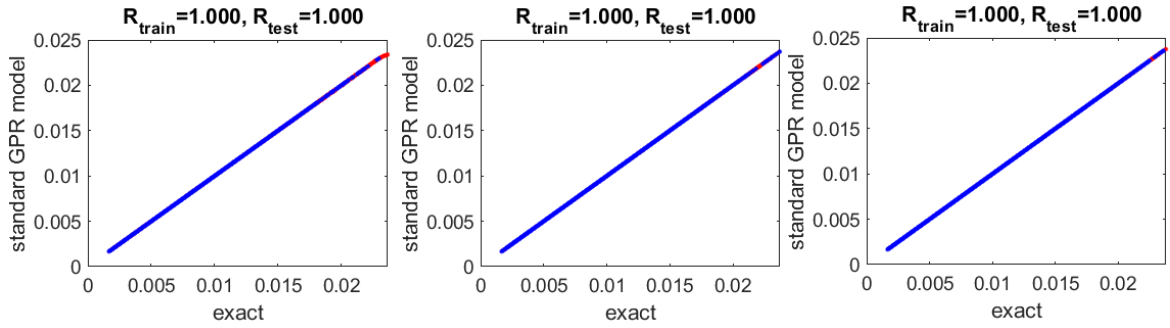

Figure 3. Correlation plots between partially spatially averaged Kohn-Sham KED values $\overline{\tau_{KS}}$ ("exact", in a.u.) and predicted $\overline{\tau_{KS}}(\bm{x})$ for different numbers of train points (left to right: 2000, 5000, 10000). Blue: train points, red: test points. Where test points are not visible, they visually overall with the training points.

To illustrate the effect of $n_{ave}$, we plot in Figure 4 the minimum and maximum values of the MRE of $B'$ over all three materials achieved with different random point selections for different $n_{ave}$, for the case of 5000 training points. The case $n_{ave} = 0$ corresponds to no averaging. Already $n_{ave} = 5$ leads to a noticeable improvement of the quality of the model, and $n_{ave} = 10$ is sufficient to significantly palliate the problem of sampling very unevenly distributed data with relatively few training points. $n_{ave} = 10$ corresponds in our case to moving averaging with a window width of about 2.2 Å, which is sufficient to simultaneously smoothen the data distribution and thereby significantly facilitate machine learning, and to preserve spatial variation of the features and of the KED.

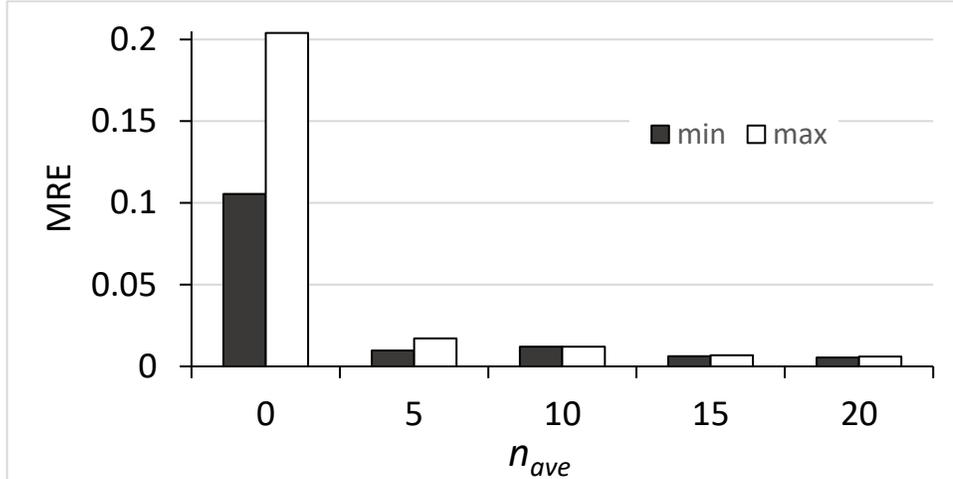

Figure 4. Minimum and maximum values of the MRE of $B'$ over all three materials achieved with different random point selections for different $n_{ave}$, for the case of 5000 training points. The case $n_{ave} = 0$ corresponds to no averaging.

## 4 Conclusion

The development of kinetic energy *density* functionals is one of the major approaches to the construction of kinetic energy functionals for orbital-free DFT. KEDFs have the advantage of being well-suited for achieving near linear scaling, and they are conceptually easy to understand. Machine learning of KED as a function of density-based descriptors intrinsically requires working with large datasets – samples in real space of the KED and density-dependent features, which are large even for a single material at a single geometry. Such samples need to be collected at different geometries (to provide good energy-geometry dependence) and for different materials (for the KEDF to possess portability), which further increases the amount of data. The values of the KED and of the features, which in general include powers and derivatives of the density and in this work are terms of the 4$^{th}$ order gradient expansion and of the product of the density and the Kohn-Sham effective potential, are very unevenly distributed. This complicates the sampling of the feature space, requires more training data, causes overfitting, and ultimately lowers the quality of a ML KED model.

We have shown that it is advantageous to work with partially spatially averaged (smoothed) KED and density-dependent features. Smoothing on a scale of about 2 Å makes sampling much easier while retaining the spatial dependence necessary to machine-learn a KED. When fitting averaged KEDs of Al, Si, and Mg simultaneously with GPR, reliable models can be built from as few as 2000 data (out of a total dataset of more than a half a million data points including data for all three materials at equilibrium volume and with compressive and tensile strains), with no significant variation due to different random draws of the training points. This is in spite of the fact that when fitting original data, even 10000 training points are not sufficient to suppress significant variations in the results due to this randomness. Because features that enter a non-linear GPR kernel, the effect of their averaging is non-linear and is different from averaging the prediction as a linear operation which is always operated in GPR simply by virtue of a finite width of the kernel.

*While it is trivial that fitting smoothed data is easier, what is important in our results is not the improvement of the (smoothed) KED error (which is natural with smoothing) but the resulting improvement in the computed kinetic energy, and more importantly, improvement in energy-volume dependence* which is critical for structure optimization and phononic properties calculations. We obtained an accuracy of the quantity *B'*, that mimics the bulk modulus, on the order of 1% (compared to the parent KS DFT calculation) for all three materials simultaneously with not more than 5000 training points. Without the averaging, even with 10000 training points, *B'* was accurate to 5-10% and with significant variations due to random selection of the training set.

Our results show that data distribution is a significant issue in machine learning of KED. Despite the growing literature on machine learning for KEF construction, the data aspect thereof remains understudied. Addressing this issue, and the associated issue of the density of sampling, is both necessary and fruitful for building more accurate KEDF for OF-DFT. We hope that the present results will spur further work in this direction.

## 5   Supplementary material

The supplementary material includes data distribution graphs.

## 6   Acknowledgements

This work was supported by JST-Mirai Program Grant Number JPMJMI22H1, Japan.

## 7 Data availability statement

Due to their large size, the data used in this study are available from the authors at reasonable request.

[23] S. Das, M. Iyer, and V. Gavini, "Real-space formulation of orbital-free density functional theory using finite-element discretization: The case for Al, Mg, and Al-Mg intermetallics," Phys. Rev. B **92**(1), 014104 (2015).

[24] B.G. Radhakrishnan, and V. Gavini, in *Recent Progress in Orbital-Free Density Functional Theory* (WORLD SCIENTIFIC, 2012), pp. 147–163.

[25] B. Radhakrishnan, and V. Gavini, "Effect of cell size on the energetics of vacancies in aluminum studied via orbital-free density functional theory," Phys. Rev. B **82**(9), 094117 (2010).

[26] V. Gavini, K. Bhattacharya, and M. Ortiz, "Quasi-continuum orbital-free density-functional theory: A route to multi-million atom non-periodic DFT calculation," J. Mech. Phys. Solids **55**(4), 697–718 (2007).

[27] K.M. Carling, and E.A. Carter, "Orbital-free density functional theory calculations of the properties of Al, Mg and Al–Mg crystalline phases," Modelling Simul. Mater. Sci. Eng. **11**(3), 339 (2003).

[28] F. Legrain, and S. Manzhos, "Highly accurate local pseudopotentials of Li, Na, and Mg for orbital free density functional theory," Chem. Phys. Lett. **622**, 99–103 (2015).

[29] C. Huang, and E.A. Carter, "Transferable local pseudopotentials for magnesium, aluminum and silicon," Phys. Chem. Chem. Phys. **10**(47), 7109–7120 (2008).

[30] Q. Liu, D. Lu, and M. Chen, "Structure and dynamics of warm dense aluminum: a molecular dynamics study with density functional theory and deep potential," J. Phys.: Condens. Matter **32**(14), 144002 (2020).

[31] H. Zhuang, M. Chen, and E.A. Carter, "Elastic and Thermodynamic Properties of Complex Mg-Al Intermetallic Compounds via Orbital-Free Density Functional Theory," Phys. Rev. Applied **5**(6), 064021 (2016).

[32] J.-D. Chai, V.L. Lignères, G. Ho, E.A. Carter, and J.D. Weeks, "Orbital-free density functional theory: Linear scaling methods for kinetic potentials, and applications to solid Al and Si," Chem. Phys. Lett. **473**(4), 263–267 (2009).

[33] J.C. Snyder, M. Rupp, K. Hansen, L. Blooston, K.-R. Müller, and K. Burke, "Orbital-free bond breaking via machine learning," J. Chem. Phys. **139**(22), 224104 (2013).
Page 23 of 26